\documentclass[prx,noshowpacs,print,twocolumn]{revtex4}
\usepackage{amssymb}
\usepackage{amsmath}
\usepackage{graphicx}
\usepackage{dcolumn}
\usepackage{bm}
\usepackage{txfonts}
\usepackage{appendix}
\usepackage[dvips]{color}
\usepackage[colorlinks,linkcolor=red,citecolor=blue]{hyperref}
\begin{document}

\title{Even-odd dependent optical transitions of zigzag monolayer black phosphorus nanoribbons}
\author{Pu Liu$^1$, Xianzhe Zhu$^1$, Xiaoying Zhou$^1$\footnote{xiaoyingzhou@hunnu.edu.cn}, Benliang Zhou$^1$, Wenhu Liao$^2$,
Guanghui Zhou$^1$\footnote{ghzhou@hunnu.edu.cn}, and Kai Chang$^3$\footnote{kchang@semi.ac.cn} }

\affiliation{$^1$Department of Physics, Key Laboratory for
Low-Dimensional Structures and Quantum Manipulation (Ministry of
Education) and Synergetic Innovation Center for Quantum Effects and
Applications of Hunan, Hunan Normal University, Changsha 410081,
China}

\affiliation{$^2$Department of Physics and Key Laboratory of Mineral
Cleaner Production and Exploit of Green Functional Materials in
Hunan Province, Jishou University, Jishou 416000, China}

\affiliation{$^3$SKLSM, Institute of Semiconductors, Chinese Academy of Sciences, P.O. Box 912, Beijing 100083, China}

\begin{abstract}
We analytically study the electronic structures and optical
properties of zigzag-edged black phosphorene nanoribbons (ZPNRs)
utilizing the tight-binding (TB) Hamiltonian and Kubo formula. By
solving the discrete Schordinger equation directly, we obtain the
energy spectra and wavefunctions for a $N$-ZPNR with $N$
number of transverse zigzag atomic chains, and classify the
eigenstates according to the lattice symmetry. We then obtain the
optical transition selection rule of ZPNRs based on the symmetry
analysis and the analytical expressions of the optical
transition matrix elements. Under an incident light
linearly-polarized along the ribbon, importantly, we find that the optical transition selection
rule for the $N$-ZPNR with even- or odd-$N$ is qualitatively
different. In specification, for even-$N$ ZPNRs the inter- (intra-)
band selection rule is $\Delta n=$odd (even), since the parity of
the wavefunction corresponding to the $n$th subband in the
conduction (valence) band is $(-1)^{n}[(-1)^{(n+1)}]$ due to the
presence of the $C_{2x}$ symmetry. In contrast, all optical
transitions are possible among all subbands due to the absence of
the $C_{2x}$ symmetry. Our findings provide a further understanding
on the electronic states and optical properties of ZPNRs, which are
useful in the explanation of the optical experiment data on ZPNR
samples.
\end{abstract}

\maketitle

\section{Introduction}
Black phosphorus (BP) is a layered material similar to graphite with
the atomic layers coupled by van der Waals interactions
\cite{YBZhang,Ye,StevenP,FengnianXia}. Few-layer
\cite{YBZhang,Ye,StevenP,FengnianXia,Buscema} and monolayer
\cite{Ye,Andres,Lu,xmwang} BP (termed as phosphorene) have been
fabricated experimentally, attracted intensive attentions due to
their unique electronic properties and potential applications in
nanoelectronics \cite{Xiling,Gomez,YBZhangPL}. Unlike graphene, BP
is a semiconductor possessing a direct band gap ranging from 0.3 eV
to 1.8 eV depending on the thicknesses of BP samples
\cite{xmwang,YBZhangPL}. The field-effect-transistor (FET) based on
phosphorene is found to have an on/off ratio of 10$^{3}$ and a
carrier mobility of 800 cm$^{2}$/V$\cdot $s \cite{Sherman}. Sizable
band gap and relatively high mobility in phosphorene bridge the gap
between graphene and transition metal dichalcogenides (TMDs), which
are important for electronics and optoelectronics
\cite{Xiling,Gomez,YBZhangPL}. Inside phosphorene, phosphorus atoms
are covalently bonded with three adjacent atoms to form a puckered
honeycomb structure due to the $sp^{3}$ hybridization \cite{Rodin}.
Arising from the low symmetric and high anisotropic structure, BP
exhibits strongly anisotropic electrical
\cite{Ye,xyzhou,xyzhouoptic,rzhang,xyzhougfactor}, optical
\cite{Tony2,Tran,YBZhangPL,xmwang} and transport \cite{Zhenhua}
properties.

The band structure of 2D phosphorene can be well described by a four
band tight-binding (TB) model \cite{Rudenko,Rudenkogap}. Tailoring
it into 1D nanoribbons offer us a way to tune its electronic and
optical properties due to the quantum confinement and unique edge
effects \cite{Tran,Carvalhopnr,hanxy,ezawa,Taghizadeh,Zahra}. The
band structure of phosphorene nanoribbons (PNRs) depends on the edge
configurations \cite{Tran,Carvalhopnr,hanxy,ezawa,Taghizadeh,Zahra}.
The armchair-edged PNRs (APNRs) are semiconductors with direct band
gap sensitively depending on the ribbon width with scaling law of
$1/N^2$ \cite{Tran,Taghizadeh,Zahra}, while the bare zigzag-edged
PNRs (ZPNRs) are metallic regardless of their ribbon width due to
the quasi-flat edge states \cite{Carvalhopnr,ezawa}. In bare ZPNRs,
the edge states are entirely detached from the bulk bands and
localized at the boundaries. These edge states result in a
relatively large density of states near the Fermi energy
\cite{ezawa,Longlong Zhang}. Further, the band structure of ZPNRs
can be effectively modified by tensile strain \cite{hanxy} or
electric field \cite{hanxy,ezawa,HGuo}. Very recently,
few-layer ZPNRs are successfully synthesized in recent experiments
\cite{Paulmd,NakanishiAyumi}. Up to date, various interesting
properties have been predicted for ZPNRs, including those related to
transverse electric field controlled FET \cite{ezawa}, room
temperature magnetism and half metal phase \cite{wucj,yujia,Reny},
strain induced topological phase transition \cite{Sisakhtet}, and
symmetry dependent response to perpendicular electric fields
\cite{Zhoubl}, etc.

On the other hand, although there are already many research works on
2D phosphorene and its 1D ribbons, the analytical calculation on the
band structure of ZPNR is still lacking. Most of the previous works
on this issue are based on the first-principles calculation
\cite{Tran,Carvalhopnr,hanxy,HGuo} or numerical diagonalization
utilizing the TB model \cite{ezawa,Taghizadeh}. As well, there is
also less attention has been paid to the optical property of ZPNR
\cite{Zahra,Sima}, and particularly the optical transition selection
rule in relation to the lattice symmetry and wavefunction parity are
not fully understood. Optical spectrum measurements are fundamental
approach to detect and understand the crystal band structure, which
have been successfully performed for 2D phosphorene
\cite{YBZhangPL}. To this end, in this work we theoretically
investigate the optical properties of ZPNRs based on the TB model
and the Kubo formula. By solving the discrete Schordinger equation
analytically, we obtain the electronic structures of ZPNRs and
classify their eigenstates according to the crystal symmetry. We
then obtain the optical transition selection rules of ZPNRs directly
based on the symmetry analysis and the analytical expressions of the
optical transition matrix elements. When the incident light is
polarized along the ribbons (see Fig. 1), interestingly, we find
that the optical selection rules change significantly for a $N$-ZPNR
with even- or odd-$N$. In particular, for even-$N$ ZPNRs the
electronic wavefunction parity of the $n$th subband in the
conduction (valence) band is $(-1)^{n}[(-1)^{(n+1)}]$ due to the
$C_{2x}$ symmetry, and therefore their inter- (intra-) band
selection rule is $\Delta n$=$n-n'$=odd (even). For odd-$N$ ZPNRs
without $C_{2x}$ symmetry, in contrast, the optical transitions are
all possible among subbands. Further, the edge states of both even-
and odd-$N$ ZPNRs play an important role in the optical absorption.
Moreover, impurities or external electric field can break the
$C_{2x}$ symmetry of even-$N$ ZPNRs, which consequently enhances the
optical absorption.

The paper is organized as follows. Sec. II mainly presents
the analytical result. We first repeat the numerical diagonalization
procedure to obtain the band structure for the system, and the
detailed analytical calculations on the band structure with
particularity of approaching accurate edge bands are followed. Then
the wavefunctions, the joint density of states and the optical
conductivity for ZPNRs are expressed. In Sec. III, we present some
numerical examples and discussions on the band structure and optical
absorptions of the ZPNRs. Finally, we summarize our results in Sec.
IV.

\begin{figure}
\includegraphics[width=0.48\textwidth, bb=0 44 827 590]{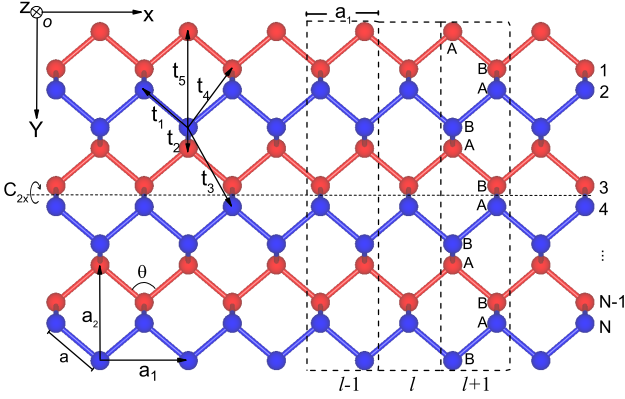}
\caption{Top view of a even-$N$ ZPNR, where the red (blue) spheres
represent phosphorous atoms in the upper (lower) sub-layer with
primitive vectors $|\bm{a}_1|=3.32$ \r{A} and $|\bm{a}_2|=4.38$
\r{A} of 2D phosphorene. The bond length between two adjacent atoms
is $a$=2.207 \r{A} with bond angle $\theta$=96.79$^{\circ}$. The
(black) dashed-line rectangles are suppercells adopted here for TB
diagonalizing calculation. }
\end{figure}

\section{Electronic structure and optical properties}
\subsection{Numerical diagonalization on Hamiltonian}
The puckered honeycomb structure of phosphorene is shown in
Fig. 1, where the red and blue dots represent phosphorous atoms
in different sub-layers. There are four atoms in one unit cell with
the primitive vectors $|\bm{a}_1|=3.32$ \r{A} and $|\bm{a}_2|=4.38$
\r{A}. The bond length between two adjacent atoms is $a$=2.207 \r{A}
with bond angle $\theta$=96.79$^{\circ}$ \cite{Rudenko}. Tailoring
phosphorene into 1D nanoribbons along the zigzag direction leading
to ZPNRs. The length of the bond connecting different sub-layers is
2.207 $\r{A}$ and the layer spacing is $l=2.14$ \r{A}. The integers
1, 2, $\cdots$ $N$ describe the number of the zigzag atomic chains
of a ZPNR along its transversal direction. In the TB framework
\cite{Rudenko,Rudenkogap}, the Hamiltonian of a phosphorene in the
presence of in-plane transverse and of-plane vertical electric
fields as well as impurities can be generically written as
\begin{equation}
H=\sum\limits_{<i,j>}t_{ij}c_{i}^{\dag
}c_{j}+\sum_{i}(\frac{1}{2}eE_{v}l\mu_{i}+eE_{t}y_i+U_i)c_{i}^{\dag }c_{i},
\end{equation}
where the summation $\langle i,j\rangle$ runs over all neighboring
atomic sites with hopping integrals $t_{ij}$, and $c^{\dag}_{i}$
($c_j$) is the creation (annihilation) operator for atom site
$i$($j$). It has been shown that five hopping parameters (see Fig.
1) are enough to describe the electronic band structure of
phosphorene \cite{Rudenko} with hopping energies $t_1$=$-$1.22 eV,
$t_2$=3.665 eV, $t_3$=$-$0.205 eV, $t_4$=$-$0.105 eV, and
$t_5$=$-$0.055 eV. A uniform vertical electric field $E_v$ will
result in a staggered potential $elE_v$ between the upper
($\mu_i$=1) and lower ($\mu_i=-1$) sublayers due to the puckered
structure \cite{Zhoubl, R.Ma}. Applying a transverse electric field
$E_t$ will shift the on-site energy to $eE_{t}y_i$ with $y_i$ the
atom coordination in the $y$-direction, and $U_i$ is the impurity
potential.

For a $N$-ZPNR with the number of zigzag chains $N$ across the
width, by applying the Bloch's theorem the TB Hamiltonian in the
momentum space is \cite{Datta}
\begin{equation}
H=H_{00}+H_{01} e^{ik_x a_1}+H_{01}^{\dag}e^{-ik_x a_1},
\end{equation}
where Hamiltonian $H_{00}$ ($H_{01}$) describes the intra (inter)-
supercell [see the (black) dashed-line rectangles in Fig. 1]
interactions, $k_x$ is the wavevector along the $x$-direction. In
our calculation, we accordingly choose the basis ordered as
($|1A\rangle,|1B\rangle,|2A\rangle,|2B\rangle,\cdots |mA\rangle,
|mB\rangle,\cdots |NA\rangle,|NB\rangle)^T$ to write done $H_{00}$
and $H_{01}$ in the form of ($2N\times2N$) matrix for the super cell
adopted. Then, we can obtain the energy spectrum $E_{n,k_x}$ and the
corresponding wavefunction $|n,k_x\rangle$ for the system by
numerical diagonalization. In real space, the wavefucntion can be
formly expressed as
\begin{equation}
\psi_{n,k_x}(\bm{r})=\sum_{m=1}^{N}\sum_{i=A,B}\frac{e^{ik_xx_{m,i}}}{\sqrt{L_x}}
\frac{c_{m,i}}{\sqrt{2N\pi}\alpha}e^{-\frac{(\bm{r}-\bm{R}_{m,i})^2}{\alpha}},
\end{equation}
where $\bm{r}$=$(x,y)$ is the electron coordination,
$\bm{R}_{m,i}$=$(x_{m,i} ,y_{m,i})$ is the atomic position vector,
$\{c_{m,i}\}^T$= $[c_{1A},c_{1B},c_{2A},c_{2B},\cdots
c_{NA},c_{NB}]^T$ is the eigenvector of the Hamiltonian matrix in
Eq. (2) with the transpose operator $T$, and $\alpha$ is a Guass
broadening parameter. Up to now, the band structure of ZPNRs is well
understood by the first-principles calculations
\cite{Tran,Carvalhopnr,hanxy,HGuo} or numerical TB calculations
\cite{ezawa,Taghizadeh}. For comparison here, the band structure of
our numerical diagonalization for a 10-ZPNR is shown by the (black)
solid lines in Fig. 2(a), which is in good agreement with the
existed results \cite{ezawa,Taghizadeh}. We note that there is a
little difference compared with that of the first-principles
calculation \cite{Tran,Carvalhopnr,hanxy,HGuo} due to the relaxation
of the edge atoms. Considering the limitation of the first-principles calculation, the TB model can be applied to study the ZPNRs with large widths. More importantly, we give the analytical solutions for electronic states and optical transitions in the ZPNRs with arbitrary widths. In comparison with the previous
numerical calculations \cite{Tran,
ezawa,Taghizadeh,Carvalhopnr,hanxy,HGuo}, the analytical results are
more convenient to do further understand in the electronic property
of ZPNRs, i.e., identifying the subband symmetry property and
calculating the optical absorption. Hereafter, we will present the
analytical calculations on the band structure of ZPNRs in the next
subsection.

\subsection{Analytical calculation on electronic structure}
In this subsection, we firstly outline a scheme to obtain the
analytical energy spectrum for ZPNRs by solving the TB model
directly. According to the TB approximation, the discrete
Schordinger equation for a $N$-ZPNR is
\begin{equation}
\begin{aligned}
E\phi_{A}(m) &  =t_{1}g_{k}\phi_{B}(m)+t_{2}\phi_{B}(m-1)+t_{3}g_{k}\phi_{B}(m-2)\\
&+t_{4}g_{k}[\phi_{A}(m-1)+\phi_{A}(m+1)]+t_{5}\phi_{B}(m+1),\\
E\phi_{B}(m) &  =t_{1}g_{k}\phi_{A}(m)+t_{2}\phi_{A}(m+1)+t_{3}g_{k}\phi_{A}(m+2)\\
&+t_{4}g_{k}[\phi_{B}(m-1)+\phi_{B}(m+1)]+t_{5}\phi_{A}(m-1).%
\end{aligned}
\end{equation}
where $g_{k}$=$2\text{cos}(k_{x}a_{1}/2)$, $\phi_{A/B}(m)$ is the
wavefuntion of the $m$th A/B atom, and the site index $m=0, 1, 2,
\cdots N+1$. Since the $0$B and $(N+1)$A sites are missing,
we naturally have the hard-wall boundary condition for ZPNRs as
\begin{equation}
\begin{aligned}
\phi_{B}(0)=\phi_{A}(N+1)=0.
\end{aligned}
\end{equation}
According to the Bloch theorem, the generic solutions for
$\phi_{A}(m)$ and $\phi_{B}(m)$ can be written as
\begin{equation}
\begin{aligned}
\phi_{A}(m)=Ae^{ipm}+Be^{-ipm},\;\; \phi_{B}(m)=Ce^{ipm}+De^{-ipm},
\end{aligned}
\end{equation}
where $A$, $B$, $C$ and $D$ are arbitrary coefficients and $p$ the
wavenumber in the transverse direction, which can be defined by the
Schordinger equation combined with the boundary condition.
Substituting Eq. (6) into (5), the wavefunction $\phi_{A/B}(m)$ can
be simplified as
\begin{equation}
\begin{aligned}
\phi_{A}(m)  &  =A(e^{ipm}-z^{2}e^{-ipm})=A\varphi_{A}(p,m),\\
\phi_{B}(m)  &  =C(e^{ipm}-e^{-ipm})=C\varphi_{B}(p,m),
\end{aligned}
\end{equation}
where $z=e^{ip(N+1)}$. Meanwhile, substituting Eq. (7) into (4), we
obtain a matrix equation
\begin{equation}
M\left(
\begin{array}
[c]{c}%
A\\
C
\end{array}
\right)=0,
\end{equation}
where M is a 2$\times$2 matrix with elements
\begin{equation*}
\begin{aligned}
M_{11}=&E\varphi_{A}(p,m)-g_{k}t_{4}[\varphi_{A}(p,m-1)+\varphi_{A}(p,m+1)],\\
M_{12}=&-[t_{1}g_{k}\varphi_{B}(p,m)+t_{2}\varphi_{B}(p,m-1)\\
&+t_{5}\varphi_{B}(p,m+1)+t_{3}g_{k}\varphi_{B}(p,m-2)],\\
M_{21}=&-[t_{1}g_{k}\varphi_{A}(p,m)+t_{2}\varphi_{A}(p,m+1)\\
&+t_{5}\varphi_{A}(p,m-1)+t_{3}g_{k}\varphi_{A}(p,m+2)],\\
M_{22}=&E\varphi_{B}(p,m)-g_{k}t_{4}[\varphi_{B}(p,m-1)+\varphi_{B}(p,m+1)].
\end{aligned}
\end{equation*}

The condition for nontrivial solutions of $A$ and $C$ in Eq. (8),
namely $[A, C]^T\neq0$, is det($M$)=0. However, it is worth to note
that the solutions of $p=0$ and $\pm\pi$ should be excluded as
unphysical results because these values of $p$ yield
$\phi_{A/B}(m)$=0 [see Eq. (7)] for arbitrary $m$. In other words,
electrons are absent in the system in these cases, which is
unphysical. Therefore, we should find solutions that satisfy
det($M$)=0 for arbitrary $m$ except $p=0$, $\pm\pi$. After some
arithmetic, we find that the equation det($M$)=0 yields the
following equation
\begin{equation}
ve^{i2pm}+we^{-i2pm}+\xi=0,
\end{equation}
where $v$, $w$ and $\xi$ are functions of $E$, $k_x$ and $p$.
Generally, Eq. (9) should be valid for arbitrary $m$. Thus, the two
coefficients ($v$ and $w$) of $e^{\pm i2pm}$ and the constant term
$\xi$ should be zero. We then obtain the energy spectrum for ZPNR as
\begin{equation}
\begin{aligned}
E=2g_{k}t_{4}\cos(p)\pm\left\vert t_{1}g_{k}+t_{2}e^{ip}+t_{5}e^{-ip}%
+t_{3}g_{k}e^{2ip}\right\vert,
\end{aligned}
\end{equation}
where $\pm$ represent the conduction and valence bands,
respectively.

On the other hand, from $\xi$=0, we find a transcendental equation
for the transverse wavevector $p$ which can be determined by
\begin{equation}
\begin{aligned}
F(p,N,k)=&t_{1}g_{k}\sin[p(N+1)]+t_{2}\sin(pN)\\
&+t_{3}g_{k}\sin[p(N-1)]+t_{5}\sin[p(N+2)]=0.
\end{aligned}
\end{equation}
This equation implies that the transverse wavenumber $p=p(k_{x},N)$
depends not only on the ribbon width $N$ but also on the
longitudinal wavenumber $k_{x}$. Obviously, we have
$F(p,N,k)=-F(-p,N,k)$, which means that Eq. (11) defines the same
subbands for $p\in(-\pi,0)$ and $p\in(0,\pi)$. Hence, we can simply
find the solutions of $p$ from Eq. (11) in the later interval. If
$t_1$=$t_2$ and $t_3=t_5=0$, Eq. (11) reduces to the transcendental
equation for a zigzag-edged graphene nanoribbon (ZGNR) case
\cite{Katsunori,Saroka}. Similar to that in a $N$-ZGNR, there are
only $N$-$1$ nonequivalent solutions of Eq. (11) for $p\in(0,\pi)$,
which defines $2N$-$2$ subbands, namely the bulk states of a ZPNR.

Notably, the two edge states are naturally missing in the scheme
here since the transverse wavevector are purely imaginary as is
described by Eq. (6). But fortunately we can restored them by
setting $p=i\beta$ and do the same procedure above to obtain the
eigenenergy and the corresponding transcendental equation. In this
case, the eigenenergy in Eq. (10) can be rewritten as
\begin{equation}
\begin{aligned}
E&=g_{k}t_{4}(e^{\beta}+e^{-\beta})\pm\sqrt{f(\beta)f(-\beta)},
\end{aligned}
\end{equation}
where
$f(\beta)=t_{1}g_{k}+t_{2}e^{\beta}+t_{3}g_{k}e^{2\beta}+t_{5}e^{-\beta}$,
with the corresponding transcendental equation expressed by
\begin{equation}
\begin{aligned}
G(\beta,N,k)&=t_{1}g_{k}\sinh[\beta (N+1)]+t_{2}\sinh(\beta N)\\
&+t_{3}g_{k}\sinh[\beta (N-1)]+t_{5}\sinh[\beta(N+2)]=0,
\end{aligned}
\end{equation}
where $\sinh(x)$ is the hyperbolic sine function. Obviously, we have
$G(\beta,N,k)$=$-G(-\beta,N,k)$, which means we only need to find
the solution of $\beta$ for $\beta>0$ case.

Therefore, according to Eq. (10), we can obtain the band structure
of ZPNRs by this analytical approach. We present an example of bulk
band structure in Fig. 2(a), where the (red) dashed lines describe
the analytical result for 10-ZPNR, which exactly matches our
previous numerical one [see the (black) solid lines] starting from
Eq. (2). In addition, the edge bands of different 10-, 15- and
30-ZPNR are also shown in Figs. 2(b-d), where the (black) solid and
(blue) dash-dotted lines represent the numerical and analytical
results, respectively. Unfortunately, we can see that the analytical
results for edge states given by Eq. (12) are not in consistent with
the numerical ones. This discrepancy was also revealed in a recent
work \cite{M. Aminic}. We think that the discrepancy mainly
originates from the hopping links $t_3$, $t_4$ and $t_5$, with which
their hopping distances are beyond a zigzag chain (see Fig. 1). This
makes the discrete Schordinger equation (4) is invalid for the edge
atoms, namely $m$ equal to 1 or $N$. To resolve this problem, one
solution is choosing four atoms to write down Eq. (4) and double the
number of the boundary condition (5). But this method will enlarge
the dimensions of matrix $M$ unavoidably and make the problem to be
quite complicate and difficult to solve.

Hereby, we propose an efficient solution to eliminate this
discrepancy by simply adding a correction term. Generally, the two
edge states can be described by a 2$\times$2 matrix Hamiltonian as
\begin{equation}
H_{edge}=\left(
\begin{array}
[c]{cc}%
h_0 & h_{c}\\
h_{c}^{\ast} & h_0
\end{array}
\right),
\end{equation}
where $h_0$ describes the two degenerate edge states when the ribbon
width $N$ is large, and $h_c$ describes the coupling between the
edge states for small $N$. Based on this argument, according to Eq.
(12), we have $h_0=2t_{4}g_{k}\sinh(\beta)$ and
$h_c=\sqrt{f(\beta)f(-\beta)}$. The band structure of the edge
states in 10-, 15- and 30-ZPNR given by Eq. (12) are presented by
the (blue) dash-dotted lines in Figs. 2(b-d), respectively. From
these figures, we find that $h_c$ is finite for a narrow ribbon as
shown in Fig. 2(b) for 10-ZPNR, but it vanished for the wider ones
[e.g., Fig. 2(d) for 30-ZPNR]. This means that $h_c$ is suitable to
describe the coupling between the edge states. However, there is a
observable discrepancy between the analytical results [Eq. (12)]
shown by the (blue) dash-dotted lines and the numerical ones [the
(black) solid lines] in Figs. 2(b-d). This implies that $h_0$ is
unsuitable to describe the edge states and needs a correction so as
to describe the edge bands accurately. Naturally, we can assume that
the correction term is a superposition of the energy term caused by
hopping links ($t_3$, $t_4$, $t_5$) beyond one zigzag chain, which
is expressed as
\begin{equation}
\begin{aligned}
h'=&\sum_{s=+,-}(b_3^st_3g_ke^{s2\beta}+b_4^st_4g_ke^{s\beta}+b_5^st_5e^{s\beta}),
\end{aligned}
\end{equation}
where the coefficients $b_3^+=0.01783$, $b_3^-=0.5739$,
$b_4^{\pm}=-1.419$, $b_5^+=-0.1345$ and $b_5^-=8.735$. Then the
energy of edge states can be written as
\begin{equation}
\begin{aligned}
E=h_0+h'\pm h_c.
\end{aligned}
\end{equation}
The analytical edge states expressed by Eq. (16) are also depicted
by the (red) dashed lines in Figs. 2(b-d). Comparing them with the
numerical data [the (black) solid lines], we find that they are in
excellent agreement with each other regardless of the ribbon width,
which indicates that our method is valid and reliable.
\begin{figure}
\includegraphics[width=0.43\textwidth,bb=75 6 697 525]{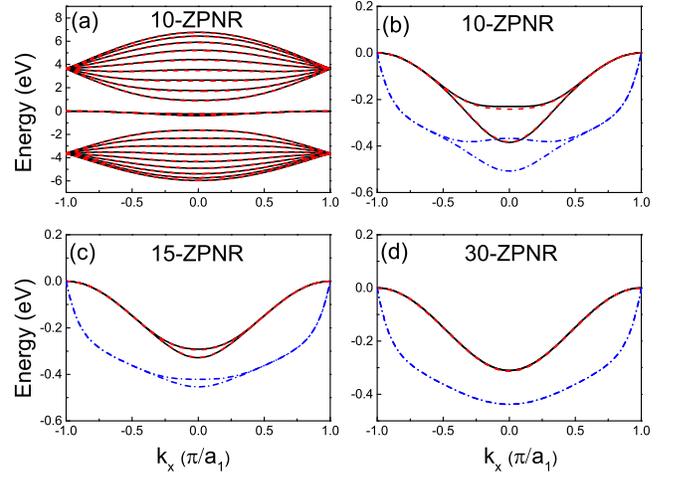}
\caption{(a) The band structure of a bare 10-ZPNR, where the
(red/black) dashed/solid lines represent the analytical/numerical
results. The scale-enlarged edge bands for (b) 10-ZPNR, (c) 15-ZPNR
and (d) 30-ZPNR with the comparison of analytical and numerical
results, where the [red (blue)] dashed (dash-dotted) lines represent
the analytical result for the edge bands with (without) the
correction term $h'$. }
\end{figure}

On the other hand, for the wavefunction (3), owing to the
translational invariance along the $x$-direction we can rewrite  it
in another generic form
\begin{eqnarray}
\psi_{n,k_x}(r)&=&\sum_{m=1}^{N}\left(
\begin{array}
[c]{cc}%
C_A\varphi_{A}(p,m)e^{ik_{x}x_{m,A}} \\
C_B\varphi_{B}(p,m)e^{ik_{x}x_{m,B}}
\end{array}
\right),
\end{eqnarray}
where $C_{A}$ and $C_{B}$ are the normalization coefficient, $N$ the
number of A and B atoms in a super-cell of a ZPNR, and
$x_{A/B,m}$ the $x$-coordination of the $m$th $A/B$ atoms. Notably,
Eq. (17) is the wavefunction for the bulk states of ZPNRs. For edge
states, we should replace the transversal wavevector $p$ by
$i\beta$.

As for the wavefunction of even-$N$ ZPNRs, we can obtain the
relation \textbf{$C$=$\mp Az$} in Eq. (7) from the parity of a ZPNR,
namely $\phi_A(N+1-m)$=$\pm \phi_B(m)$ \cite{Katsunori}. The reason
is that the wavefunction of even-$N$ ZPNRs is either symmetric or
antisymmetric which is similar to that in ZGNRs \cite{Mahdi
Moradinasab}. Specifically, combined with the translational
invariance along the $x$-direction, the wavefuntion of even-$N$
ZPNRs is specified as
\begin{eqnarray}
\psi_{n,k_x}(r)&=&\frac{C}{\sqrt{L_x}}\sum_{m=1}^{N}\left(
\begin{array}
[c]{cc}%
-s z^{-1}\varphi_{A}(p,m)e^{ik_{x}x_{m,A}} \\
\varphi_{B}(p,m)e^{ik_{x}x_{m,B}}
\end{array}
\right)\nonumber\\
&=&\frac{C'}{\sqrt{L_x}}\sum_{m=1}^{N}\left(
\begin{array}
[c]{cc}%
-s\sin[p(N+1-m)]e^{ik_{x}x_{m,A}} \\
\sin(pm)e^{ik_{x}x_{m,B}}
\end{array}
\right),
\end{eqnarray}
where $C'$=$[\sum_{m=1}^N\sin^2(pm)]^{-1}/\sqrt{2}$ is the
normalization coefficient, $s=\pm 1$ indicates the parity of the
subbands. Notably, Eq. (18) is the wavefunction for bulk states of
even-$N$ ZPNRs. For edge states, the wavefunction ($p$=$i\beta$) is
\begin{eqnarray}
\psi_{n,k_x}(r)=\frac{C_e}{\sqrt{L_x}}\sum_{m=1}^{N}\left(
\begin{array}
[c]{cc}%
-s \sinh[\beta(N+1-m)]e^{ik_{x}x_{m,A}} \\
\sinh(\beta m)e^{ik_{x}x_{m,B}}
\end{array}
\right),
\end{eqnarray}
where $C_e$=$[\sum_{m=1}^N\sinh^2(pm)]^{-1}/\sqrt{2}$ is also a
normalization coefficient. On the contrary, owing to the absence of
the $C_{2x}$ symmetry, there is no such a simple expression of
wavefunction for the odd-$N$ ZPNRs.

\subsection{Optical property and transition selection rules}
In order to detect the above calculated band structure of ZPNRs, we
study its optical response in this subsection. One useful physical
quantity to understand the optical property is the joint density of
states (JDOS) representing all possible optical transitions among
the subbands, which is generally given by
\begin{equation}
D_J(\omega)=\frac{g_s}{
L_x}\sum_{n,n',k_x}[f(E_{n,k_x})-f(E_{n',k_x})]\delta(E_{n,k_x}-E_{n',k_x}+\hbar\omega),
\end{equation}
where the sum runs over all states $|n,k_x\rangle$ and
$|n',k_x\rangle$, $g_s$ is 2 for spin degree, $L_x$ the ribbon
length, $\hbar \omega$ the photon energy, and
$f(E)=1/[\text{exp}{(E-E_F)/k_BT}+1]$ the Fermi-Dirac distribution
function with Boltzman constant $k_B$ and temperature $T$. Here, we
take a Guass broadening
$\frac{1}{\Gamma\sqrt{2\pi}}\text{exp}[-(E_{n,k_x}-E_{n',k_x}+\hbar
\omega)^2/2\Gamma^2]$ to approximate the $\delta$-function, where
$\Gamma$ is a phenomenological constant accounting for the energy
level broadening factor. Meanwhile, assuming the incident light is
polarized along the longitudinal ($x$-) direction, the optical
conductance based on the Kubo formula is given by \cite{Ando1,Ando2}
\begin{equation}
\begin{aligned}
\sigma(\omega)=&\frac{g_s \hbar e^2}{iL_x}\sum_{n,n',k_x}
\frac{[f(E_{n,k_x})-f(E_{n',k_x})]|\langle
n,k_x|v_x|n',k_x\rangle|^{2}}{(E_{n,k_x}-E_{n',k_x})(E_{n,k_x}-E_{n',k_x}+\hbar\omega+i\Gamma)},
\end{aligned}
\end{equation}
where $v_x$=$\frac{1}{i\hbar}\frac{\partial H}{\partial k_x}$ is the
velocity operator, which is valid and independent of the band
structure model, and $|n,k_x\rangle=\phi(r)\varphi(K)$ \cite{Burt} is the total
electron wavefunction in a ZPNR. Here $\phi(r)$ is the envelop
function which describes the slowly varying electron sharing
movement in the crystal, while $\varphi(K)$ is the band edge
wavefunction (BEW) connecting to the atom orbits directly describing
the fast movement in the crystal. In a ZPNR, $\varphi(K)$ is
composed by $|s\rangle$, $|p_x\rangle$, $|p_y\rangle$, and
$|p_z\rangle$ atomic-orbits with different weights \cite{Ruo-Yu
Zhang,WeifengLi}. For a linear polarized light, the optical
transition matrix elements satisfy $\langle
n,k_x|v_x|n',k_x\rangle$=$\langle
\psi_{n,k_x}|v_x|\psi_{n',k_x}\rangle\langle
\varphi_n(K)|\varphi_{n'}(K)\rangle$. Obviously,
$v_{n,n'}(k_x)$=$\langle \psi_{n,k_x}|v_x|\psi_{n',k_x}\rangle$
determines the optical transition selection rules. A zero matrix
element $v_{n,n'}(k_x)$ means a forbidden transition. The inner
product between the two BEWs is subband dependent, with which only
affects the amplitude of the optical conductance but does not change
the optical selection rules. We take the inner production around the
$\Gamma$-point
($\langle\varphi_n(\Gamma)|\varphi_{n'}(\Gamma)\rangle$) as an
approximation and treat it as a constant. This approximation has
also been used in the previous work \cite{Tony2} for 2D phosphorene.
We have omitted this constant in our calculations because the
specific expression of the BEWs in ZPNRs are currently unknown. This
approximation would not change the essential physics, i.e., the
even-odd dependent optical selection rule, reported here. Note that
in some topological none-trivial system, the dipole optical matrix
result in the winding number \cite{Likunshi,Tingcao}. But there is
no such an effect in phosphorene because it is a topologically
trivial system. The real part of $\sigma(\omega)$ indicates the
optical absorption when an laser beam incidents on the sample.
Moreover, we can obtain the dielectric function
$\varepsilon(\omega)$ from optical conductance by using
$\varepsilon(\omega)$=1+$\frac{4\pi i}{\omega}\sigma(\omega)$
\cite{Peteryu}.

\begin{figure}
\includegraphics[width=0.47\textwidth,bb=75 3 740 547]{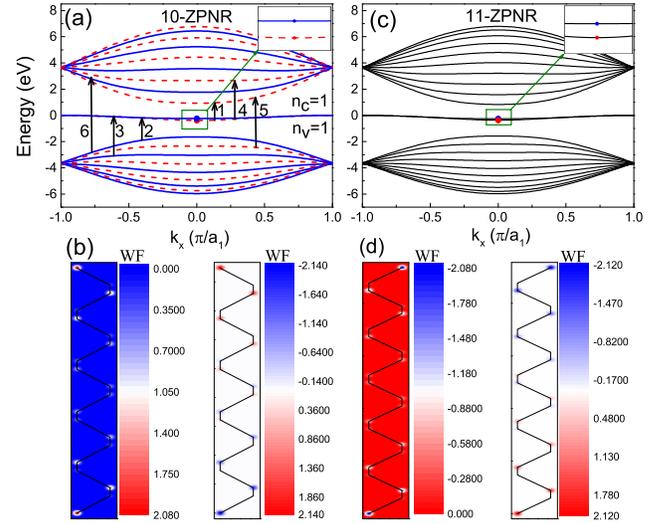}
\caption{The band structures of bare (a)10-ZPNR and (c)11-ZPNR,
where the (red/blue) dashed/solid lines represent the
symmetric/antisymmetric states. (b) and (d) show the spatial
distribution of the wave function of the subband $n_v=1$ (left
panel) and $n_c=1$ (right panel) corresponding to the states at
$k_x$=0 indicated by the red and blue dots in (a) and (c),
respectively.}
\end{figure}

In optical transition process, selection rules determined by the
matrix elements $v_{n,n'}(k_x)$ are the most important information.
The integral of the velocity matrix elements $V_{n,n'}$=$\int dk_x
|v_{n,n'}(k_x)|^2$ is proportional to the optical transition
probability between the $n$th and $n'$th subbands. Generally, the
selection rule is always constrained by the symmetry of the system.
Hence, in order to obtain a general optical selection rule for
ZPNRs, we firstly check their lattice symmetry. According to Fig. 1,
we find the lattice symmetry of a $N$-ZPNR is even-odd-$N$
dependent. In particular, the even-$N$ ones have a
$C_{2x}(x,y,z)$$\rightarrow$$(x,-y,-z)$ operator with respect to the
ribbon central axis (see the dotted horizontal line in Fig. 1). This
is equivalent to the symmetry operator $\sigma_{zx}\sigma_{xy}$,
where $\sigma_{zx}$ and $\sigma_{xy}$ are the mirror symmetry
operators corresponding to the $xoz$ and the $xoy$ planes,
respectively. However, the odd-$N$ ones do not have this symmetry.
In even-$N$ ZPNRs, the constraint on the eigenstates $\langle
x,y,z|n,k_x\rangle$ imposed by $C_{2x}$ symmetry is $C_{2x}\langle
x,y,z|n,k_x\rangle$=$\langle x,-y,-z|n,k_x\rangle$. Assuming
$\lambda$ is the eigenvalue of $C_{2x}$ operator, we have
\begin{equation}
\langle x,y,z|n,k_x\rangle=(C_{2x})^2\langle x,y,z|n,k_x\rangle=\lambda^2\langle x,y,z|n,k_x\rangle.
\end{equation}
Then we obtain $\lambda^2$=1, i.e., $\lambda$=$\pm$1, where $+/-$
means the even/odd parity provided by the $C_{2x}$ operator. This
indicates that $\langle x,y,z|n,k_x\rangle$ is either symmetric or
antisymmetric along the $y$ and $z$ directions, namely
$\langle-y,-z|n,k_x\rangle$=$\pm$$\langle y,z|n,k_x\rangle$. Thus,
we can classify the eigenstates for even-$N$ ZPNRs as even or odd
parity according to the eigenvalues of the $C_{2x}$ operator.
\begin{figure}
\includegraphics[width=0.48\textwidth,bb=18 281 609 524]{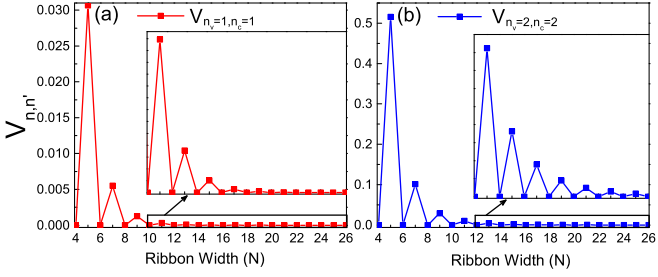}
\caption{The integral of the optical transition matrix elements
$V_{n,n'}$=$\int dk_x |v_{n,n'}(k_x)|^2$ for (a) the $n_v$=1 subband
to the $n_c$=1 one and (b) the $n_v$=2 subband to the $n_c$=2 one as
a function of the ribbon width $N$, where the insets show the amplified picture of $V_{n,n'}$ on ribbon width with large $N$.}
\end{figure}

In order to confirm the above argument on the symmetry and parity
for the systems, we present the band structure and the wavefunction
in real space of the first subband in the conduction and valence
bands for 10- and 11-ZPNR in Figs. 3(a-b) and 3(c-d), respectively.
The wavefunction corresponding to states indicated by the red (blue)
dots in Figs. 3(a) and 3(c) are shown in the left (right) panels in
Figs. 3(b) and 3(d), respectively. According to the left (right)
panel in Fig. 3(b), we find that the first subband in the conduction
(valence) band is even (odd) under $C_{2x}$ transformation. By
checking the eigenstates in other subbands, we observe that the
parity of wavefunctions varies alternatively from odd [(blue) solid
lines] to even [(red) dashed lines] with the increase of subband
index $n$. This is consistent with the previous results obtained by
the first-principles calculation \cite{Zahra,Tran}. Hence, the
parity of the subband in the conduction (valence) band is related to
its subband index via $(-1)^{n}[(-1)^{(n+1)}]$. Further, under the
$C_{2x}$ operation, the velocity operator $v_x$ is even, i.e.,
$C_{2x}:v_x\rightarrow v_x$. Hence, we obtain the condition for none
zero matrix element $v_{n,n'}(k_x)$ is that the parity of the
initial ($|n,k_x\rangle$) and final ($|n',k_x\rangle$) states are
the same. In other words, only the transitions among the states with
identical parity are allowed. This can also be verified by
calculating the optical transition matrix element. For example,
using the relation $\mathbf{v}=\frac{i}{\hbar}[\mathbf{r},H]$
combined with the wavefunction Eq. (18), the inter-band optical
transition matrix element between the bulk states is \cite{Peteryu}
\begin{equation}
\langle \psi_v |v_x|\psi_c\rangle=\frac{i}{\hbar}\langle \psi_v |Hx-xH|\psi_c\rangle,
\end{equation}
$\psi_{c/v}$ is the wavefunction in Eq. (18) or
(19). After some algebra, we have
\begin{equation}
\langle \psi_{n,kx}^v|v_x|\psi_{n',kx}^c\rangle=
\left\{
  \begin{array}{ll}
    \frac{i}{\hbar}\frac{C'^2}{L_x}(A_{t_1}+A_{t_3}+A_{t_4}),\;\; s'=s \\
    0,\;\; s'\neq s,
  \end{array}
\right.
\end{equation}
where
\begin{equation*}
\begin{aligned}
&A_{t_1}=-4i t_{1}b\sin(bk_{x})\sum_{m=1}^{N}\sin(pm)\sin[p^{\prime}(N+1-m)],\\
&A_{t_3}=-4i t_{3}b\sin(bk_{x})\sum_{m=1}^{N-2}\sin(pm)\sin[p^{\prime}(N-1-m)],\\
&A_{t_4}=8i t_{4}b\sin(bk_{x})\cos(p')\sum_{m=1}^{N}\sin(pm)\sin(p^{\prime}m).
\end{aligned}
\end{equation*}
Here $b=a_1/2$ and the detailed calculations on Eq. (24) are
presented in the Appendix A.

From Eq. (24), we can explicitly find that only the inter-band
transition between the bulks states with the same symmetry are
allowed. Using the wavefunction in Eq. (19), we can obtain the same
selection rules $s=s'$ for the transition between the edge bands as
well as that between the bulk bands and edge bands. Hence, we
conclude that only the transitions between the subbands with same
parity are allowed. Consequently, in even-$N$ ZPNRs, the inter
(intra)-band selection rule is $\Delta n$=$n-n'$=odd (even). This is
in good agreement with the above analysis based on the lattice
symmetry. It is important that although the band structure of
odd-$N$ ZPNRs is similar to that of even-$N$ ones as shown in Fig.
3(c), the optical selection rule is qualitatively different from
that for even-$N$ ZPNRs. According to the left (right) panel in Fig.
3(d), by checking the eigenstates within the whole band, we know
that there is no subband owning definite parity in 11-ZPNR due to
the absence of $C_{2x}$ symmetry. Thus, the optical transitions in
odd-$N$ ZPNRs between two arbitrary subbands are all possible. In
order to illustrate the even-odd dependent optical selection rule
more clearly, in Fig. 4 we show the integral of the optical
transition matrix elements $V_{n,n'}$ as a function of the ribbon
width $N$, where (a) for $V_{n_v=1,n_c=1}$ and (b) for
$V_{n_v=2,n_c=2}$, respectively, and the insets show the amplified picture of $V_{n,n'}$ on ribbon width with large $N$. Physically, $V_{n,n'}$ is
proportional to the optical transition probability between the $n$th
and $n'$th subbands. According to the figure, we find that the
transition probability oscillates with the ribbon width $N$ and shows
an even-odd $N$ dependent feature. The transitions between the
subband $n_v$=1 (2) and $n_c$=1 (2) are forbidden in even $N$-ZPNRs
due to the presence of the $C_{2x}$ symmetry. In contrast, the
transitions between the subband $n_v$=1 (2) and $n_c$=1 (2) are
allowed in odd $N$-ZPNRs due to the absence of the $C_{2x}$
symmetry. This even-odd dependent selection rule is also reflected
in the optical absorption spectrum which will be discussed in the
nest section.

\section{Numerical Results and Discussions}
In this section, we present some numerical examples for the optical
absorption spectrum of ZPNRs and discuss the corresponding results.
We take $N$=10 and 11 to represent the even and odd cases,
respectively, which would not qualitatively influence the results
here. The temperature is 4 K and the level broadening $\Gamma$ is 4
meV throughout the calculation unless specificated. In all following
figures, the green solid line (if available) indicates the Fermi
level.

\begin{figure}
\includegraphics[width=0.48\textwidth,bb=63 3 762 586]{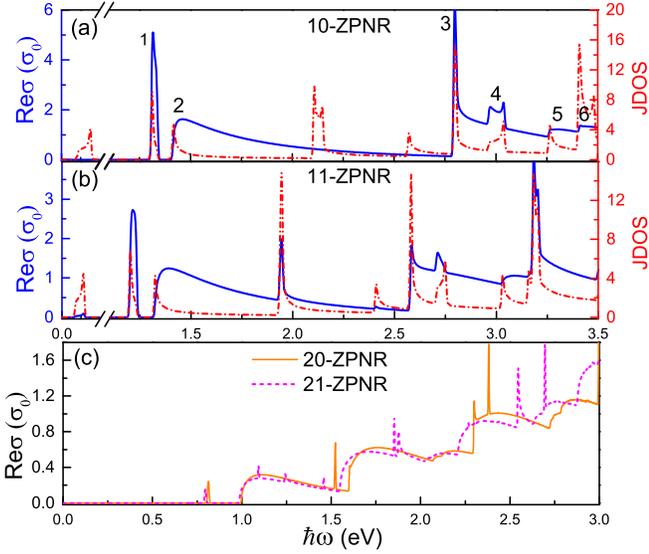}
\caption{The inter-band JDOS [(red) dash-dotted lines] and the
optical absorption [(blue) solid lines] as a function of the
incident photon energy with $\sigma_0=2e^2/h$ for (a) 10-ZPNR and (b)
11-ZPNR. The Fermi level $E _F$ is chosen as $-$0.3086 eV lying
between the two bands of edge states. The peaks (labeled by 1, 2,
$\cdots$ 6) are associated with the subband transitions illustrated
in Fig. 2(a). (c) The optical absorption spectra for 20-ZPNR
[(orange) solid line] and 21-ZPNR [(purple) dashed line].}
\end{figure}

As discussed in Sec. IIC, the inter- (intra-) band optical
transition selection rule in even-$N$ ZPNRs satisfies $\Delta
n$=$n-n'$=odd (even) due to the $C_{2x}$ symmetry. On the contrary,
the optical transitions in odd-$N$ ZPNRs between two arbitrary
subbands are all possible resulting from the $C_{2x}$ symmetry
breaking. Keeping this in mind is important to understand the
optical properties of even-$N$ ZPNRs. Fig. 5 shows the inter-band
JDOS and the optical absorption spectrum for (a) 10-ZPNR and (b)
11-ZPNR with Fermi energy $E_F=-$0.3086 eV lying between the edge
states, respectively. As shown by the (red) dash-dotted line in Fig.
5(a), we see peaks in the JDOS spectrum at different photon energy
known as van Hove singularities. The JDOS peaks range from the
mid-infrared (155-413 meV) to the visible region due to the edge
states and the quantum confinement, which is different from that of
2D phosphorene case \cite{YBZhangPL}. However, there is no optical
absorption around zero frequency, which is contradict to the fact
that ZPNRs are metallic. The reason is that the transition between
the edge states is forbidden by the $C_{2x}$ symmetry in even-$N$
ZPNRs since their parities are different from each other. Compared
with the JDOS, we find that more peaks are missing in the optical
absorption spectra Re$\sigma(\omega)$ [the (blue) solid line] due to
the optical selection rule $\Delta n$=odd arising from the $C_{2x}$
symmetry, which is similar to that in ZGNRs \cite{Han,Chung,Saroka}.
The remained optical absorption peaks (labeled by 1, 2, $\cdots$ 6)
originating from the allowed transitions contributed by subbands
with the same parity which are schematically illustrated in Fig.
3(a). In contrast, as shown in Fig. 5(b), the optical transitions
among subbands are all possible for 11-ZPNR owing to the $C_{2x}$
symmetry breaking. All the optical absorption peaks appear one to
one correspondence to the JDOS. Owing to the edge states, the
absorption peaks range from the mid-infrared to the visible
frequency, etc. However, the absorption peak in the mid-infrared
frequency (the first peak) disappears for wider ribbons as shown in
Fig. 5(c), which is different from that in ZGNRs
\cite{Han,Chung,Saroka}. The reason comes from two sides: i) the
edge states become degenerate for wider ZPNRs and ii) unlike that in
ZGNRs, the edge states of ZPNRs are slightly dispersed [see Fig.
3(a)] due to the electron-hole asymmetry. This fact means that only
two $k_x$ states contribute to the optical absorption for a certain
Fermi level, leading to zero optical conductance. Again, from Fig.
5(c), we find that there are more absorption peaks for 21-ZPNR than
that of the 20-ZPNR arising from the $C_{2x}$ symmetry breaking.
Moreover, it should be noted that there is a little discrepancy
between the JDOS peaks and the optical absorption peaks because that
the optical transition matrix element $v_{n,n'}(k_x)$ depends on the
subbands' derivatives $\partial E/\partial_{k_x}$.

\begin{figure}
\includegraphics[width=0.48\textwidth,bb=22 42 736 538]{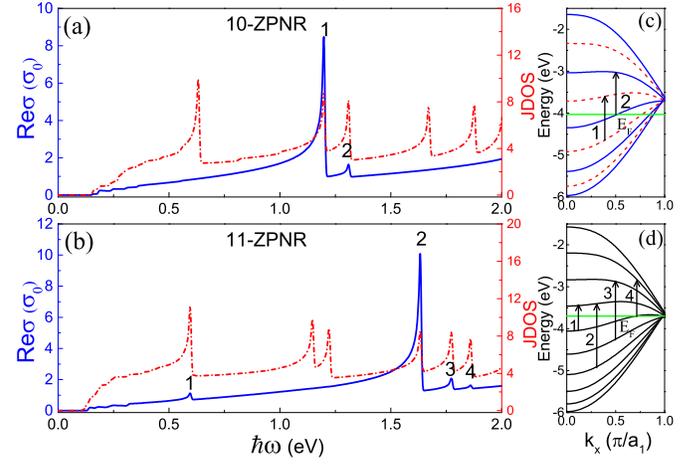}
\caption{The intra-band JDOS [(red) dash-dotted lines] and the
optical absorption [(blue) solid lines] as a function of the
incident photon energy for (a) 10-ZPNR and (b) 11-ZPNR. The band
structure and the corresponding optical transitions are shown in (c)
and (d) for 10-ZPNR ($E_F$=$-$4.0351 eV) and 11-ZPNR
($E_F$=$-$3.7241 eV), respectively.}
\end{figure}

Figure 6 shows the intra-band JDOS [(red) dash-dotted line] and
optical absorption spectrum [(blue) solid line] for (a) 10-ZPNR and
(b) 11-ZPNR with the corresponding band structures and Fermi levels
shown in (c) and (d), respectively. As depicted in Fig. 6(a), the first
JDOS peak for 10-ZPNR located at $\hbar\omega=$0.604 eV comes from
the transition between the $n_v$=6 subband and the $n_v$=5 one. But
there is no absorption peak at the same frequency because the
parities of the two subbands are different [see Fig. 6(c)] and the
transitions are forbidden by the $C_{2x}$ symmetry. In other words,
this transition violates the intra-band optical transition selection
rule ($\Delta n$=even) for even-$N$ ZPNRs. By the same token, the
second and third JDOS peaks are contributed by the transitions
between the subbands with the same parities [see Fig. 6(c)], hence
the corresponding absorption peaks appear [see the (blue) solid
line]. On the contrary, we find that the optical absorption peaks
are almost presented for 11-ZPNR [see Fig. 6(b)] arising from the
$C_{2x}$ symmetry breaking which means that the all optical
transitions are principally possible among all subbands. The
corresponding transitions are shown in Fig. 6(d). On the other hand,
some of the matrix elements $\langle n,k_x|v_x|n',k_x\rangle$ may be
tiny (weak), i.e, the $\langle 6(7),k_x|v_x|4(5),k_x\rangle$, and
the corresponding absorption peaks are missing in this case [see
Fig. 6(b)].

\begin{figure}
\includegraphics[width=0.48\textwidth,bb=20 40 735 537]{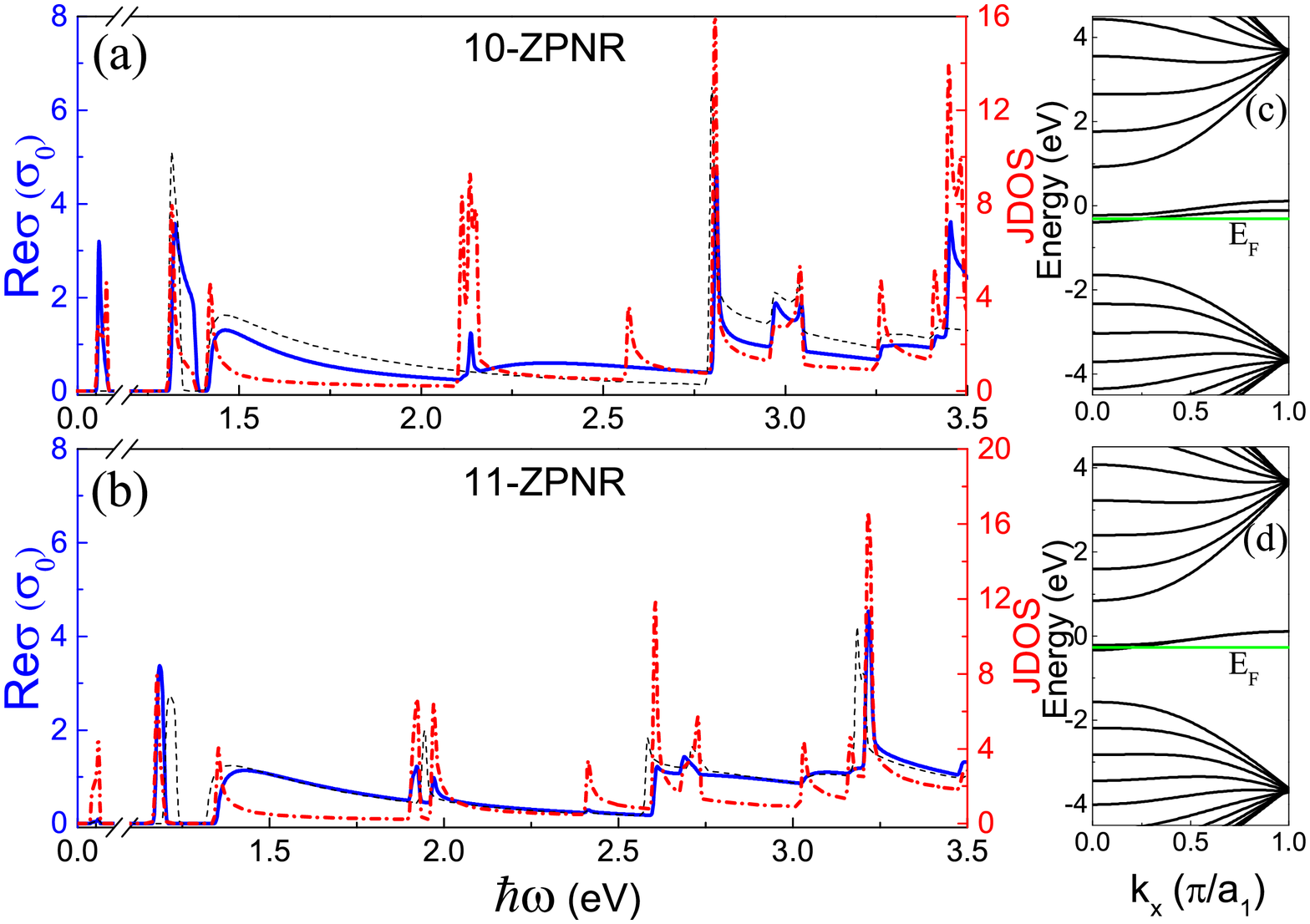}
\caption{The inter-band JDOS [(red) dash-dotted line] and optical
absorption spectrum [(blue) solid line] for (a) 10-ZPNR ($E_F$=-0.31
eV) and (b) 11-ZPNR ($E_F=-$0.2747 eV) with a uniform vertical
electric field $E_v$=0.1V/$\r{A}$, respectively, where the (black)
dash line is the absorption spectrum for bare ZPNRs. The band
structures of 10- and 11-ZPNRs under electric field are shown in (c)
and (d), respectively, where the Fermi levels for both cases are
lying between the two edge states.}
\end{figure}

Next, we turn to the effect of externally applied electric field on
the optical property of ZPNRs. Fig. 7 depicts the inter-band JDOS
[(red) dash-dotted line] and optical absorption spectra [(blue)
solid line] for (a) 10-ZPNR and (b) 11-ZPNR under a uniform vertical
electric field (VEF) with strength $E_v$=0.1V/$\r{A}$, where the
corresponding band structures with the optical transition
indications are shown in (c) and (d), respectively. The Fermi level
for both cases are lying between the edge states. In real
experiment, the VEF corresponding to the top gate or substrate
effect. It maybe be generated by using the polar semiconductors
interface \cite{Dong Zhang}. Owing to the puckered lattice structure
of ZPNRs, the band structure of the ZPNR under a VEF is even-odd
dependent \cite{Zhoubl} since the edge states of even (odd)-$N$
ribbons located on the different (same) sub-layers. The VEF opens a
gap between the two edge bands for even ribbons [see Fig. 7(c)], but
for odd ones the two edge bands are always (nearly) degenerated [see
Fig. 7(d)]. Further, the VEF also breaks the $C_{2x}$ symmetry in
even-$N$ ZPNRs. These features are also reflected in the optical
absorption spectrum. From Fig. 7(a), we find that several extra
absorption peaks [the (blue) solid line] appear due to the $C_{2x}$
symmetry breaking by the VEF compared with the bare 10-ZPNR [see the
(black) dashed line]. Especially, the first absorption peak in
mid-infrared frequency is greatly enhanced due to the degeneracy
lifting of the edge states. In comparison, as shown in Fig. 7(b),
the absorption spectrum of 11-ZPNR is slightly changed compared to
the bare case [also see the (black) dashed line] since the band
structure is nearly unaffected by the VEF. These features offer a
useful approach to identify the even-odd property of ZPNR samples by
experimentally detecting the optical absorption under VEF.

\begin{figure}
\includegraphics[width=0.48\textwidth,bb=25 12 735 514]{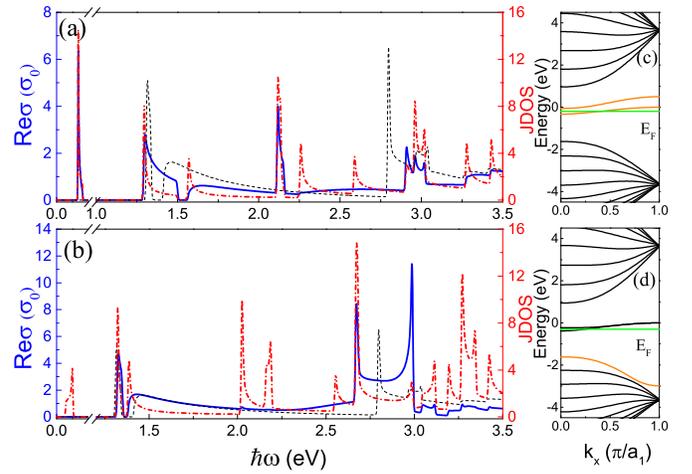}
\caption{The inter-band JDOS [(red) dash-dotted line] and optical
absorption spectrum [(blue) solid line] for 10-ZPNR with impurities
localized at (a) one edge (the 1st atomic row) with $E_F$=-0.1954 eV
and (b) the center (the 10th row) with $E_F=-$0.3058 eV,
respectively, where the impurity potential $U_i$ is 0.5 eV and 1.5
eV corresponding to (a) and (b), and the (black) dashed line
indicates the optical spectrum of a pristine 10-ZPNR. While (c) and
(d) respectively to (a) and (b) but for band structure.}
\end{figure}

Experimentally, it is difficult to avoid impurities and defects in
samples. This may consequently affect the optical properties of
ZPNRs by changing the band structure or breaking the $C_{2x}$
symmetry. Figs. 8 (c) and 8(d) show the band structure of 10-ZPNR
with impurities distributing on its one edge (the 1st atomic row)
and the center (the 10th row), respectively, where we have defined a
$N$-ZPNR with 2$N$ atomic rows. For a zigzag chain of ZPNR there are
two phosphorus atomic rows, hence a $N$-ZPNR has 2$N$ rows. We model
the impurity effect by adding a impurity potential $U_i$ to the
on-site energy of the corresponding impurity atoms, which is widely
used in previous works \cite{zouyl,L.L.Li}. As shown in Fig. 8(c)
for impurities located on the edge, we find that the nearly
degenerated edge states are separated [see the (orange) solid line]
due to the variation of the on-site energies, but the other subbands
remain unchanged. This is consistent with the result obtained by the
first-principles calculation \cite{Pooja,guocx}. On the contrary,
comparing Fig. 8(d) with Fig. 3(a), as impurities localized on the
center the subband contributed by the impurities are shifted [see
the (orange) solid line] but the other subbands remain unchanged.
This means that the impurities only have a local effect on the
electronic structure of a ZPNR. But, they will play an important
role in the optical absorption spectrum because of the lattice
symmetry breaking. Figs. 8(a) and 8(b) show the inter-band JDOS
[(red) dash-dotted line] and optical absorption spectrum [(blue)
solid line]for 10-ZPNR with impurities located at its one edge (the
1st atomic row) with $E_F$=-0.1954 eV and the center (the 10th row)
with $E_F=-$0.3058 eV, respectively. The impurities localized at the
edge break the $C_{2x}$ symmetry since the wavefunctions
corresponding to most of the subbands are partially distributed on
the edge. Hence, we observe the first and some extra optical
absorption peaks reappeared [see the (blue) solid line] compared
with the pristine 10-ZPNR shown in Fig. 8(a) [(black) dash line].
This is similar to that of the VEF effect discussed above.
Similarly, from Fig. 8(b), we also find some extra peaks when the
impurities localized at the center. However, the absorption peak at
the mid-infrared frequency (the first peak) is still missing
although the $C_{2x}$ symmetry is broken in this case. The reason is
that the edge states are mainly localized on the edge atoms, in
consequence the band structure is nearly unaffected by the
impurities localized on the center atoms. For 11-ZPNR, the optical
absorption spectrum is just slightly changed by the impurities,
hence we do not present the result here for saving space.

\begin{figure}
\includegraphics[width=0.48\textwidth,bb=25 30 746 517]{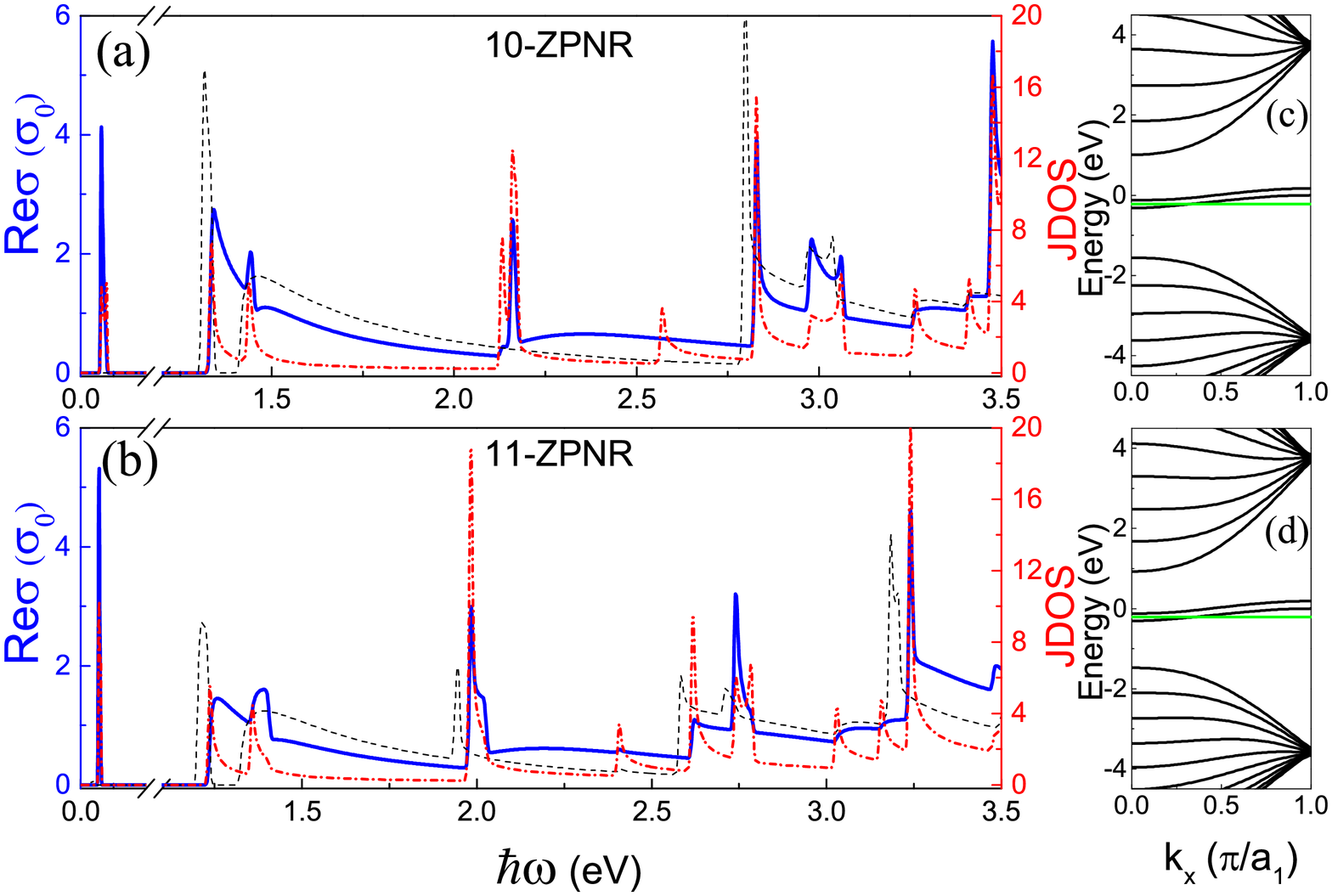}
\caption{The inter-band JDOS [(red) dash-dotted line] and optical
absorption spectrum [(blue) solid line] for (a) 10-ZPNR
($E_F=-$0.2228 eV) and (b) 11-ZPNR ($E_F=-$0.2151 eV) under a
uniform TEF with strength $E_t$=0.008 V/$\r{A}$, respectively, where
the Fermi level for both cases are lying between the two edge states
and the (black) dash line represents the absorption spectrum for
bare ZPNRs. The band structures of 10- and 11-ZPNR under TEF are
shown in (c) and (d), respectively.}
\end{figure}

Finally, a transverse electric field (TEF) can induce a Stark effect
(potential difference) arising from the finite width of ZPNRs
\cite{ezawa}, which can make a significant change of the band
structure, especially the edge bands. And this effect has been
experimentally observed for few-layer BPs \cite{BingchenDeng}. As a
result, this can also change the optical properties of ZPNRs by
breaking the $C_{2x}$ symmetry. Therefore, Fig. 9 displays the
inter-band JDOS [(red) dash-dotted line] and optical absorption
spectrum [(blue) solid line] for (a) 10-ZPNR and (b) 11-ZPNR under a
uniform TEF with strength $E_t$=0.008 V/$\r{A}$. And the
corresponding band structures are shown in Figs. 9(c) and 9(d),
respectively. As shown in the figure, we find that in the presence
of the TEF the optical absorption peaks are shifted compared to the
bare ribbons [see the (black) dashed line]. In Figs. 9(c) and 9(d),
unlike the VEF case, we can see the degeneracy of the edge states
for 10- and 11-ZPNR are both lifted by the Stark effect. Owing to
the $C_{2x}$ symmetry breaking, all possible absorption peaks in
10-ZPNR corresponding to the JDOS appear [see Fig. 9(a)], which
means that the optical absorption of 10-ZPNR is greatly enhanced,
especially the absorption peak in the mid-infrared frequency.
Further, the first absorption peak for both 10- and 11-ZPNR is
greatly enhanced due to the degeneracy lifting as show in Figs. 9(a)
and 9(b). Hence we conclude that the effect of TEF on the optical
absorption in ZPNRs is different from that of the VEF or impurity. A
TEF can induce different potentials on all atoms within a
super-cell, which leads to a global $C_{2x}$ symmetry breaking.

\section{Summary}
In summary, we have theoretically studied the electronic and optical
properties of ZPNRs under a linearly polarized light along the
longitudinal direction based on the TB Hamiltonian and Kubo formula.
We have obtained analytically the energy spectra of ZPNRs and the
optical transition selection rules based on the lattice symmetry
analysis. Owing to the $C_{2x}$ symmetry, the eigenstates of
even-$N$ ZPNRs are transversely either symmetric or antisymmetric,
which makes their optical response qualitatively different from that
of the odd-$N$ ones. In particular the inter (intra) -band selection
rule for even-$N$ ZPNRs is $\Delta n=$ odd (even) since the parity
factor of the wavefunction corresponding to the conduction (valence)
band is $(-1)^{n}[(-1)^{(n+1)}]$ (with the subband index $n$)
provided by the $C_{2x}$ symmetry. For odd-$N$ ZPNRs, however, the
all optical transitions are possible among all subbands. Further,
the edge states play an important role in the optical absorption and
are involved in many of the absorption peaks. The optical absorption
of even-$N$ ZPNRs can be enhanced by the substrate and impurity
effect as well as the transverse electric field via breaking the
$C_{2x}$ symmetry. While the optical absorption of odd-$N$ ones can
be effectively tuned by lattice defects or external electric fields.
Our findings provide a further understanding on the electronic states and
optical properties of ZPNRs, which are essential for the explanation of the optical experiment data on ZPNR
samples.

\section{Acknowledgments}
This work was supported by the National Natural Science Foundation
of China (Grant Nos. 11804092, 11774085, 61674145, 11704118,
11664010), and China Postdoctoral Science Foundation funded project
Grant No. BX20180097, and Hunan Provincial Natural Science
Foundation of China (Grant No. 2017JJ3210).

\section*{Appendix A}
\setcounter{equation}{0}
\renewcommand{\theequation}{A\arabic{equation}}
In this appendix, we calculate the optical transition matrix
elements in Eq. (24). Utilizing the relation
$\textbf{v}=\frac{i}{\hbar}[\textbf{r},H]$ combined the wavefunction in Eq. (18), the
optical matrix element can be written as \cite{Peteryu}
\begin{equation}
\langle \psi_v |v_x|\psi_c\rangle=\frac{i}{\hbar}\langle \psi_c |Hx-xH|\psi_v\rangle,
\end{equation}
$\psi_{c/v}$ is the wavefunction in Eq. (18) or
(19). According to Eq. (18), the transition matrix element between
the bulk states is
\begin{widetext}
\begin{equation}
\begin{aligned}
\langle \psi_{n,k_x}^v|v_x|\psi_{n',k_x}^c\rangle=&\frac{C^{\prime 2}}{L_{x}}\frac{i}{\hbar}\sum_{m=1}^{N}\sum_{n=1}^{N}\Bigg\{(x_{A,n}-x_{A,m}%
)e^{-ik_{x}x_{A,m}}e^{ik_{x}x_{A,n}%
}ss^{\prime}\sin[p(N+1-m)]\sin[p^{\prime}(N+1-n)]\left\langle A_{m}\right\vert
H\left\vert A_{n}\right\rangle \\
&  +(x_{B,n}-x_{B,m})e^{-ik_{x}%
x_{B,m}}e^{ik_{x}x_{B,n}}\sin(pm)\sin(p^{\prime}n)\left\langle B_{m}%
\right\vert H\left\vert B_{n}\right\rangle \\
&  -s(x_{B,n}-x_{A,m})e^{-ik_{x}%
x_{A,m}}e^{ik_{x}x_{B,n}}\sin[p(N+1-m)]\sin(p^{\prime}n)\left\langle
A_{m}\right\vert H\left\vert B_{n}\right\rangle \\
&  -s^{\prime}(x_{A,n}-x_{B,m})e^{-ik_{x}x_{B,m}}e^{ik_{x}x_{A,n}}\sin(pm)\sin[p^{\prime}%
(N+1-n)]\left\langle B_{m}\right\vert H\left\vert A_{n}\right\rangle \Bigg\},\\
\end{aligned}
\end{equation}
where $m(n)$ is the atom site index, and $s(s')$ indicates the parity of the subbands. There are five hoppings, including
$\left\langle A_{m}\right\vert H\left\vert B_{n}\right\rangle =$ $\left\langle
B_{m}\right\vert H\left\vert A_{n}\right\rangle =t_{1}$ for $n=m$,
$\left\langle A_{m}\right\vert H\left\vert B_{n}\right\rangle =$ $\left\langle
B_{m}\right\vert H\left\vert A_{n}\right\rangle =t_{2}$ for $n=m\pm1$,
$\left\langle A_{m}\right\vert H\left\vert B_{n}\right\rangle =$ $\left\langle
B_{m}\right\vert H\left\vert A_{n}\right\rangle =t_{3}$ for $n=m\pm2$,
$\left\langle A_{m}\right\vert H\left\vert B_{n}\right\rangle =$ $\left\langle
B_{m}\right\vert H\left\vert A_{n}\right\rangle =t_{4}$ for $n=m\pm1$, and
$\left\langle A_{m}\right\vert H\left\vert B_{n}\right\rangle =$ $\left\langle
B_{m}\right\vert H\left\vert A_{n}\right\rangle =t_{5}$ for $n=m\pm1$.
Then, Eq. (A2) can be written as
\begin{equation}
\begin{aligned}
\langle \psi_{n,k_x}^v|v_x|\psi_{n',k_x}^c\rangle &=\frac{i}{\hbar}\frac{C'^2}{L_{x}}(A_{t_1}+A_{t_3}+A_{t_4}),
\end{aligned}
\end{equation}
where $A_{t_1}$, $A_{t_3}$ and $A_{t_4}$ represent the term of
transition matrix related to the hopping $t_1$, $t_3$ and $t_4$, and
the corresponding term can be written as
\begin{equation}
\begin{aligned}
A_{t_1}=&-\sum_{m=1}^{N}s\{t_{1}be^{ik_{x}b}\sin[p(N+1-m)]\sin(p^{\prime}
m)-t_{1}be^{-ik_{x}b}\sin[p(N+1-m)]\sin(p^{\prime} m)\}\\
&  +s^{\prime}\{t_{1}be^{ik_{x}b}\sin(pm)\sin[p^{\prime
}(N+1-m)]-t_{1}be^{-ik_{x}b}\sin(pm)\sin[p^{\prime
}(N+1-m)]\}\\
=&-2it_{1}b\sin(bk_{x})\sum_{m=1}^{N}\{s\sin[p(N+1-m)]\sin(p^{\prime}m)+s^{\prime
}\sin(pm)\sin[p^{\prime}(N+1-m)]\},
\end{aligned}
\end{equation}
where the sum of $m$ runs from $1$ to $N$. Defining, $n=N+1-m$, we
find $n$ also runs from $1$ to $N$ when $m\epsilon[1,N]$. Applying
the summation transform $n=N+1-m$, $A_{t_1}$ can be rewritten as
\begin{equation}
\begin{aligned}
A_{t_1}=&-2it_{1}b\sin(bk_{x})\{\sum_{n=1}^{N}s\sin(pn)\sin[p^{\prime
}(N+1-n)]+\sum_{m=1}^{N}s^{\prime}\sin(pm)\sin[p^{\prime}(N+1-m)]\}\\
=&\left\{
\begin{array}
[c]{c}%
-4i t_{1}b\sin(bk_{x})\sum_{m=1}^{N}\sin(pm)\sin[p^{\prime
}(N+1-m)],\qquad s=s^{\prime}\\
\qquad \qquad 0,\qquad \qquad \qquad \qquad \qquad \qquad \qquad  \qquad s\neq s^{\prime}.%
\end{array}
\right.
\end{aligned}
\end{equation}
Meanwhile, the term $A_{t_3}$ is
\begin{equation}
\begin{aligned}
A_{t_3}=&-\sum_{m=3}^{N}s\{be^{ik_{x}b}\sin[p(N+1-m)]\sin[p^{\prime}%
(m-2)]t_{3}-be^{-ik_{x}b}\sin[p(N+1-m)]\sin[p^{\prime
}(m-2)]t_{3}\}\\
&-\sum_{m=1}^{N-2}s^{\prime}\{be^{ik_{x}b}\sin(pm)\sin[p^{\prime}(N+1-m-2)]t_{3}%
-be^{-ik_{x}b}\sin(pm)\sin[p^{\prime}(N+1-m-2)]t_{3}\}\\
=&-2ib\sin(bk_{x})t_{3}\{\sum_{m=3}^{N}s\sin[p(N+1-m)]\sin[p^{\prime}%
(m-2)]+\sum_{m=1}^{N-2}s^{\prime}\sin(pm)\sin[p^{\prime}(N+1-m-2)]\},
\end{aligned}
\end{equation}
in this case, the atoms at the edges should be excluded because the
hopping links of $t_3$ is beyond one zigzag chain. Applying similar
summation transform in $A_{t_1}$ to $A_{t_3}$, we have
\begin{equation}
\begin{aligned}
A_{t_3}=&-2ib\sin(bk_{x})t_{3}\{\sum_{n=1}^{N-2}s\sin(pn)\sin[p^{\prime
}(N+1-n-2)]+\sum_{m=1}^{N-2}s^{\prime}\sin(pm)\sin[p^{\prime}(N+1-m-2)]\}\\
=&\left\{
\begin{array}
[c]{c}%
-4i t_{3}b\sin(bk_{x})\sum_{m=1}^{N-2}\sin(pm)\sin[p^{\prime
}(N-1-m)],\qquad s=s^{\prime}\\
\qquad \qquad 0,\qquad \qquad \qquad \qquad \qquad \qquad \qquad \qquad s\neq s^{\prime}.%
\end{array}
\right.
\end{aligned}
\end{equation}
Finally, the $A_{t_4}$ term is
\begin{equation}
\begin{aligned}
 A_{t_4}=&\sum_{m=1}^{N}\{ss^{\prime}be^{ik_{x}b}\sin[p(N+1-m)]\sin[p^{\prime
}(N+1-m+1)]t_{4}+ss^{\prime}be^{ik_{x}b}\sin[p(N+1-m)]\sin[p^{\prime
}(N+1-m-1)]t_{4}\\
&-ss^{\prime}be^{-ik_{x}b}\sin[p(N+1-m)]\sin[p^{\prime
}(N+1-m+1)]t_{4}-ss^{\prime}be^{-ik_{x}b}\sin[p(N+1-m)]\sin[p^{\prime
}(N+1-m-1)]t_{4}\\
&  +be^{ik_{x}b}\sin(pm)\sin[p^{\prime}(m-1)]t_{4}%
+be^{ik_{x}b}\sin(pm)\sin[p^{\prime}(m+1)]t_{4} \\ &-be^{-ik_{x}b}\sin(pm)\sin[p^{\prime}(m-1)]t_{4}%
-be^{-ik_{x}b}\sin(pm)\sin[p^{\prime}(m+1)]t_{4}\\
=&\sum_{m=1}^{N}4i t_{4}b\sin(bk_{x})\cos
p'\{ss^{\prime}\sin[p(N+1-m)]\sin [p^{\prime}(N+1-m)]+\sin(pm)\sin
(p^{\prime}m)\}.
\end{aligned}
\end{equation}
Here, we have used the relation $\sin(x)+\sin(y)=2\sin[(x+y)/2]\cos[(x-y)/2]$ to simplify it. Similarly, replacing all the summation index $N+1-m$ with $n$, we obtain
\begin{equation}
\begin{aligned}
A_{t_4}=&\sum_{n=1}^{N}4i ss^{\prime}t_{4}b\sin(bk_{x})\sin(pn)\sin
(p^{\prime}n)\cos p'+\sum_{m=1}^{N}4i t_{4}b\sin(bk_{x})\sin(pm)\sin
(p^{\prime}m)\cos p'\\
=&\left\{
\begin{array}
[c]{c}%
8i t_{4}b\sin(bk_{x})\cos p'\sum_{m=1}^{N}\sin(pm)\sin(p^{\prime
}m),\qquad s=s^{\prime}\\
\qquad \qquad 0,\qquad \qquad \qquad \qquad \qquad \qquad      \qquad s\neq s^{\prime}.%
\end{array}
\right.
\end{aligned}
\end{equation}
Therefore, the transition matrix elements is
\begin{equation}
\langle \psi_{n,k_x}^v|v_x|\psi_{n',k_x}^c\rangle=
\left\{
  \begin{array}{ll}
    \frac{C'^2}{L_x}\frac{i}{\hbar}(A_{t_1}+A_{t_3}+A_{t_4}),\;\; s'=s \\
    \qquad  0, \qquad \qquad\;\; s'\neq s,
  \end{array}
\right.
\end{equation}
where
\begin{equation*}
\begin{aligned}
&A_{t_1}=-4i t_{1}b\sin(bk_{x})\sum_{m=1}^{N}\sin(pm)\sin[p^{\prime}(N+1-m)],\\
&A_{t_3}=-4i t_{3}b\sin(bk_{x})\sum_{m=1}^{N-2}\sin(pm)\sin[p^{\prime}(N-1-m)],\\
&A_{t_4}=8i t_{4}b\sin(bk_{x})\cos(p')\sum_{m=1}^{N}\sin(pm)\sin(p^{\prime}m).
\end{aligned}
\end{equation*}
From Eq. (A10), we can explicitly find that only the inter band transition between the bulks states with the same symmetry are allowed. Using the wavefunction in Eq. (19), we can obtain the same selection rules $s=s'$ for the transition between the edge bands as well as the bulk bands to the edge bands. Hence, in even-$N$ ZPNRs, we conclude that only the transitions between the subbands with same parity are allowed.
\end{widetext}


\begin{references}
\bibitem{YBZhang}Likai Li, Yijun Yu, Guojun Ye, Qingqin Ge, Xuedong Ou, Hua Wu, Donglai Feng, Xian Hui Chen, and Yuanbo Zhang,
\textcolor{blue}{Nat. Nanotech. {\bf 9}, 372 (2014)}.

\bibitem{Ye}Han Liu, Adam T. Neal, Zhen Zhu, Zhe Luo, Xianfan Xu, David Tom\'{a}nek, and Peide D. Ye, \textcolor{blue}{Acs Nano. {\bf 8}, 4033 (2014)}.

\bibitem{FengnianXia}Fengnian Xia, Han Wang, and Yichen Jia, \textcolor{blue}{Nat. Commun. {\bf 5}, 4458 (2014)}.

\bibitem{StevenP}Steven P. Koenig, Rostislav A. Doganov, Hennrik Schmidt, A. H. Castro Neto, and Barbaros \"{O}zyilmaz,
\textcolor{blue}{Appl. Phys. Lett. {\bf 104}, 103106 (2014)}.

\bibitem{Buscema}Michele Buscema, Dirk J. Groenendijk, Sofya I. Blanter, Gary A. Steele, Herre S. J. van der Zant, and Andres Castellanos-Gomez,
\textcolor{blue}{Nano Lett. {\bf 14}, 3347 (2014)}.

\bibitem{Andres}Andres Castellanos-Gomez, Leonardo Vicarelli, Elsa Prada, Joshua O Island, K L Narasimha-Acharya, Sofya I Blanter, Dirk J Groenendijk,
Michele Buscema, Gary A Steele, J V Alvarez, Henny W Zandbergen, J J Palacios, and Herre S J van der Zant, \textcolor{blue}{2D Mater. {\bf 1}, 025001 (2014)}.

\bibitem{Lu}Wanglin Lu, Haiyan Nan, Jinhua Hong, Yuming Chen, Chen Zhu, Zheng Liang, Xiangyang Ma, Zhenhua Ni, Chuanhong Jin, and Ze Zhang,
\textcolor{blue}{Nano. Res. {\bf 7}, 853 (2014)}.

\bibitem{xmwang}Xiaomu Wang, Aaron M. Jones, Kyle L. Seyler, Vy Tran, Yichen Jia, Huan Zhao, Han Wang, Li Yang, Xiaodong Xu, and Fengnian Xia,
\textcolor{blue}{Nat. Nanotech. {\bf 10}, 517 (2015)}.

\bibitem{Xiling}Xi Ling, Han Wang, Shengxi Huang, Fengnian Xia, and Mildred S. Dresselhaus, \textcolor{blue}{PNAS {\bf 112}, 4523 (2015)}.

\bibitem{Gomez}Andres Castellanos-Gomez, \textcolor{blue}{J. Phys. Chem. Lett. {\bf 6}, 4280 (2015)}.

\bibitem{YBZhangPL}Likai Li, Jonghwan Kim, Chenhao Jin, Guojun Ye, Diana Y. Qiu, Felipe H. da Jornada, Zhiwen Shi, Long Chen, Zuocheng Zhang, Fangyuan Yang,
Kenji Watanabe, Takashi Taniguchi, Wencai Ren, Steven G. Louie, Xianhui Chen, Yuanbo Zhang, and Feng Wang, \textcolor{blue}{Nat. Nanotech. {\bf 12}, 21 (2017)}.

\bibitem{Sherman}Sherman Jun Rong Tan, Ibrahim Abdelwahab, Leiqiang Chu, Sock Mui Poh, Yanpeng Liu, Jiong Lu, Wei Chen, and Kian Ping Loh,
\textcolor{blue}{Adv. Mater. {\bf 30}, 1704619 (2018)}.



\bibitem{Rodin}A. S. Rodin, A. Carvalho, and A. H. Castro Neto, \textcolor{blue}{Phys. Rev. Lett. {\bf 112}, 176801 (2014)}.

\bibitem{xyzhou}X.Y. Zhou, R. Zhang, J. P. Sun, Y. L Zou, D. Zhang, W. K. Lou, F. Cheng, G. H. Zhou, F. Zhai, and Kai Chang,
\textcolor{blue}{Sci. Rep. {\bf 5}, 12295 (2015)}.

\bibitem{xyzhouoptic}Xiaoying Zhou, Wen-Kai Lou, Feng Zhai, and Kai Chang, \textcolor{blue}{Phys. Rev. B {\bf 92}, 165405 (2015)}.

\bibitem{rzhang}R. Zhang, X. Y. Zhou, D. Zhang, W. K. Lou, F. Zhai, and K. Chang, \textcolor{blue}{2D Mater. 2, 045012 (2015)}.

\bibitem{xyzhougfactor}Xiaoying Zhou, Wen-Kai Lou, Dong Zhang, Fang Cheng, Guanghui Zhou, and Kai Chang, \textcolor{blue}{Phys. Rev. B {\bf 95}, 045408 (2017)}.

\bibitem{Tony2}Tony Low, A. S. Rodin, A. Carvalho, Yongjin Jiang, Han Wang, Fengnian Xia, and A. H. Castro Neto,
\textcolor{blue}{Phys. Rev. B {\bf 90}, 075434 (2014)}.

\bibitem{Tran}Vy Tran and Li Yang, \textcolor{blue}{Phys. Rev. B {\bf 89}, 245407 (2014)}.

\bibitem{Zhenhua}Rui Zhang, Zhenhua Wu, X. J. Li, and Kai Chang, \textcolor{blue}{Phys. Rev. B {\bf 95}, 125418 (2017)}.

\bibitem{Rudenko}A. N. Rudenko and M. I. Katsnelson, \textcolor{blue}{Phys. Rev. B {\bf 89}, 201408(R) (2014)}.

\bibitem{Rudenkogap}A. N. Rudenko, Shengjun Yuan, and M. I. Katsnelson, \textcolor{blue}{Phys. Rev. B {\bf 92}, 085419 (2015)}.

\bibitem{Carvalhopnr}A. Carvalho, A. S. Rodin, and A. H. Castro Neto, \textcolor{blue}{Europhys. Lett. {\bf 108} 47005 (2014)}.

\bibitem{hanxy}Xiaoyu Han, Henry Morgan Stewart, Stephen A. Shevlin, C. Richard A. Catlow, and Zheng Xiao Guo,
\textcolor{blue}{Nano Lett. {\bf 14}, 4607 (2014)}.

\bibitem{ezawa}M. Ezawa, \textcolor{blue}{New J. Phys. {\bf 16}, 115004 (2014)}.

\bibitem{Taghizadeh}Esmaeil Taghizadeh Sisakht, Mohammad H. Zare, and Farhad Fazileh, \textcolor{blue}{Phys. Rev. B {\bf 91}, 085409 (2015)}.

\bibitem{Zahra}Nourbakhsh Zahra and Asgari Reza, \textcolor{blue}{Phys. Rev. B {\bf 94}, 035437 (2016)}.

\bibitem{Longlong Zhang}Longlong Zhang and Yuying Hao, \textcolor{blue}{Sci. Rep. {\bf 8}, 6089 (2018)}.

\bibitem{HGuo}H. Guo, N. Lu, J. Dai, X. Wu, and X. C. Zeng, \textcolor{blue}{J. Phy. Chem. C {\bf 118}, 14051 (2014)}.

\bibitem{Paulmd}Paul Masih Das, Gopinath Danda, Andrew Cupo, William M. Parkin, Liangbo Liang, Neerav Kharche, Xi Ling, Shengxi Huang,
Mildred S. Dresselhaus, Vincent Meunier, and Marija Drndi\'{c}, \textcolor{blue}{ACS Nano, {\bf 10}, 5687 (2016)}.

\bibitem{NakanishiAyumi}Yudai Nakanishi, Ayumi Ishi, Chika Ohata, David Soriano, Ryo Iwaki, Kyoko Nomura, Miki Hasegawa, Taketomo Nakamura,
Shingo Katsumoto, Stephan Roche, and Junji Haruyama, \textcolor{blue}{Nano Res. {\bf 10}, 718 (2017)}.

\bibitem{yujia}Zhili Zhu, Chong Li, Weiyang Yu, Dahu Chang, Qiang Sun, and Yu Jia, \textcolor{blue}{Appl. Phys. Lett. {\bf 105}, 113105 (2014)}.

\bibitem{wucj}Guang Yang, Shenglong Xu, Wei Zhang, Tianxing Ma, and Congjun Wu, \textcolor{blue}{Phys. Rev. B {\bf 94}, 075106 (2016)}.

\bibitem{Reny}Yi Ren, Fang Cheng, Z. H. Zhang, and Guanghui Zhou, \textcolor{blue}{Sci. Rep. {\bf 8}, 2932 (2018)}.

\bibitem{Sisakhtet}E. Taghizadeh Sisakht, F. Fazileh, M. H. Zare, M. Zarenia, and F. M. Peeters, \textcolor{blue}{Phys. Rev. B {\bf 94}, 085417 (2016)}.

\bibitem{Zhoubl}Benliang Zhou, Benhu Zhou, Xiaoying Zhou, and Guanghui Zhou, \textcolor{blue}{J. Phys. D: Appl. Phys. {\bf 50}, 045106 (2017)}.

\bibitem{Sima}Sima Shekarforoush, Daryoush Shiri and Farhad Khoeini, arXiv: 1802.02065.

\bibitem{R.Ma}R. Ma, H. Geng, W. Y. Deng, M. N. Chen, L. Sheng and D. Y. Xing, \textcolor{blue}{Phys. Rev. B {\bf 94}, 125410 (2016)}.

\bibitem{Datta}Datta S, \emph{Quantum Transport-Atom to Transistor}, Cambridge University Press, (2005).

\bibitem{Katsunori}Katsunori Wakabayashi, Ken-ichi Sasaki, Takeshi Nakanishi, and Toshiaki Enoki,
\textcolor{blue}{Sci. Technol. Adv. Mater. {\bf 11}, 054504 (2010)}.


\bibitem{Saroka}V. A. Saroka, M. V. Shuba, and M. E. Portnoi, \textcolor{blue}{Phys. Rev. B {\bf 95}, 155438 (2017)}.

\bibitem{M. Aminic} M. Aminic and M. Soltani, arXiv: 1810.03042.

\bibitem{Mahdi Moradinasab}Mahdi Moradinasab, Hamed Nematian, Mahdi Pourfath, Morteza Fathipour, and Hans Kosina, \textcolor{blue}{J. Appl. Phys. {\bf 111}, 074318 (2012)}.

\bibitem{Ando1}Tsuneya Ando, and Yasutada Uemura, \textcolor{blue}{J. Phys. Soc. Jpn. {\bf 36}, 959 (1974)}.

\bibitem{Ando2}Mikito Koshino and Tsuneya Ando, \textcolor{blue}{Phys. Rev. B {\bf 77}, 115313 (2008)}.

\bibitem{Burt}M G Burt, \textcolor{blue}{J. Phys.: Condens. Matter {\bf 4}, 6651 (1992)}. 

\bibitem{Ruo-Yu Zhang}Ruo-Yu Zhang, Ji-Ming Zheng, and Zhen-Yi Jiang, \textcolor{blue}{Chin. Phys. Lett. {\bf 35}, 017302 (2018)}.

\bibitem{WeifengLi}Weifeng Li, Gang Zhang, and Yong-Wei Zhang, \textcolor{blue}{J. Phys. Chem. C {\bf 118}, 22368 (2014)}.

\bibitem{Likunshi}Li-kun Shi, Kai Chang, and Chang-Pu Sun, arXiv:1601.04722.

\bibitem{Tingcao}Ting Cao, Meng Wu, and Steven G. Louie, \textcolor{blue}{Phys. Rev. Lett. {\bf 120}, 087402 (2018).}


\bibitem{Peteryu}Peter Y. Yu, and Manuel Cardona, \emph{Fundamentals of Semiconductors Physics and Materials Properties 4ed}, Springer (2010).




\bibitem{Han}Han Hsu and L. E. Reichl, \textcolor{blue}{Phys. Rev. B {\bf 76}, 045418 (2007)}.

\bibitem{Chung}H. C. Chung, M. H. Lee, C. P. Chang, and M. F. Lin, \textcolor{blue}{Opt. Express {\bf 19}, 23350 (2011)}.

\bibitem{Dong Zhang}Dong Zhang, Wenkai Lou, Maosheng Miao, Shou-cheng Zhang, and Kai Chang, \textcolor{blue}{Phys. Rev. Lett. {\bf 111}, 156402 (2013)}.

\bibitem{zouyl}Yong-Lian Zou, Juntao Song, Chunxu Bai, and Kai Chang, \textcolor{blue}{Phys. Rev. B {\bf 94}, 035431 (2016)}.


\bibitem{L.L.Li}L. L. Li and F. M. Peeter, \textcolor{blue}{Phys. Rev. B {\bf 97}, 075414 (2018)}.

\bibitem{Pooja}Pooja Srivastava, K. P. S. S. Hembram, Hiroshi Mizuseki, Kwang-Ryeol Lee, Sang Soo Han, and Seungchul Kim,
\textcolor{blue}{J. Phys. Chem. C {\bf 119}, 6530 (2015)}.

\bibitem{guocx}Caixia Guo, Congxin Xia, Lizhen Fang, Tianxing Wang and Yufang Liu, \textcolor{blue}{Phys. Chem. Chem. Phys. {\bf 18}, 25869 (2016)}.

\bibitem{BingchenDeng}Bingchen Deng, Vy Tran, Yujun Xie, Hao Jiang, Cheng Li, Qiushi Guo, Xiaomu Wang, He Tian, Steven J. Koester, Han Wang, Judy J. Cha,
Qiangfei Xia, Li Yang, and Fengnian Xia, \textcolor{blue}{Nat. Commun. {\bf 8}, 14474 (2017)}.



\end{references}
\end{document}